\documentclass[conference]{IEEEtran}
\IEEEoverridecommandlockouts
\usepackage{cite}
\usepackage{amsmath,amssymb,amsfonts}
\usepackage{algorithmic}
\usepackage{graphicx}
\usepackage{textcomp}
\usepackage{xcolor}
\usepackage{hyperref}
\usepackage{physics}
\usepackage{subcaption}
\usepackage{svg}
\usepackage{enumitem}
\setlist[enumerate]{itemsep=5pt, topsep=5pt}
\setlist[itemize]{itemsep=5pt, topsep=5pt}

\newcommand{\oursoftware}{\texttt{scarab}}

\usepackage{fancyhdr}

\def\BibTeX{{\rm B\kern-.05em{\sc i\kern-.025em b}\kern-.08em
    T\kern-.1667em\lower.7ex\hbox{E}\kern-.125emX}}

\makeatletter
\renewcommand{\IEEEauthorrefmark}[1]{\textsuperscript{#1}}
\makeatother

\begin{document}

\title{Software for Creating Scalable Benchmarks from Quantum Algorithms}

\author{
  \IEEEauthorblockN{
    Noah Siekierski,\IEEEauthorrefmark{1} 
    Stefan Seritan,\IEEEauthorrefmark{1}
    Neer Patel,\IEEEauthorrefmark{2}
    Siyuan Niu,\IEEEauthorrefmark{2}
    Thomas Lubinski,\IEEEauthorrefmark{3,4}
    Timothy Proctor,\IEEEauthorrefmark{1}
  }
\\

\IEEEauthorblockA{\IEEEauthorrefmark{1}\textit{Quantum Performance Laboratory, Sandia National Laboratories, Livermore, CA 94550, USA}}
\IEEEauthorblockA{\IEEEauthorrefmark{2}\textit{University of Central Florida, Orlando, FL 32816, USA}}
\IEEEauthorblockA{\IEEEauthorrefmark{3}\textit{QED-C Technical Advisory Committee -- Standards, Arlington, VA 22209, USA}}
\IEEEauthorblockA{\IEEEauthorrefmark{4}\textit{Quantum Circuits Inc, New Haven, CT 06511, USA}}
}

\maketitle

\begin{abstract}
    Creating scalable, reliable, and well-motivated benchmarks for quantum computers is challenging: straightforward approaches to benchmarking suffer from exponential scaling, are insensitive to important errors, or use poorly-motivated performance metrics. Furthermore, curated benchmarking suites cannot include every interesting quantum circuit or algorithm, which necessitates a tool that enables the easy creation of new benchmarks. In this work, we introduce a software tool for \textit{creating} scalable and reliable benchmarks that measure a well-motivated performance metric (process fidelity) from \textit{user-chosen} quantum circuits and algorithms. Our software, called \oursoftware, enables the creation of efficient and robust benchmarks even from circuits containing thousands or millions of qubits, by employing efficient fidelity estimation techniques, including mirror circuit fidelity estimation and subcircuit volumetric benchmarking. \oursoftware~provides a simple interface that enables the creation of reliable benchmarks by users who are not experts in the theory of quantum computer benchmarking or noise. We demonstrate the flexibility and power of \oursoftware~by using it to turn existing inefficient benchmarks into efficient benchmarks, to create benchmarks that interrogate hardware and algorithmic trade-offs in Hamiltonian simulation, to quantify the in-situ efficacy of approximate circuit compilation,  and to create benchmarks that use subcircuits to measure progress towards executing a circuit of interest.
\end{abstract}

%\begin{IEEEkeywords}
%Quantum Computing, Benchmarks, Benchmarking, Algorithms,  Application Benchmarks
%\end{IEEEkeywords}

%====================================================

%%% Footer with title
\pagestyle{fancy}

\renewcommand{\headrulewidth}{0.0pt}
\lhead{}
\rhead{\thepage}

\renewcommand{\footrulewidth}{0.4pt}
\cfoot{}

\lfoot{Software for Creating Scalable Benchmarks from Quantum Algorithms}

\rfoot{\today}

\section{Introduction}
\label{sec:introduction}
Over the past decade, quantum computing hardware has progressed from few-qubit physics experiments~\cite{Barends2014-ya} to commercially-available machines with dozens to hundreds of qubits~\cite{Arute2019-mk, Bluvstein2023-dp, Kim2023-si, Moses2023-do, Chen2023-la,google_quantum_ai_and_collaborators_quantum_2025}. These rapid advances have been accompanied by increasing interest in benchmarking the performance of these devices~\cite{Proctor2025-cd}. The most well-developed and widely-used benchmarks for quantum computers, such as the quantum volume benchmark~\cite{Cross2019-ku} and randomized benchmarking (RB)~\cite{Emerson2005-fd, Emerson2007-am, Knill2008-jf, Magesan2011-hc, Proctor2019-gf, Hines2023-tz, McKay2023-bx, Proctor2022-yl, Hines2023-vq, Combes2017-kr, Helsen2019-cp, Helsen2022-yp,  Magesan2012-dz}, quantify a device's performance on random circuits. However, there has been increasing interest in directly measuring the performance of contemporary quantum computing systems on algorithms and applications~\cite{Proctor2025-cd}, as demonstrated by an ever-expanding array of application-based benchmarking suites~\cite{Chen2022-dm, Linke2017-mr, Wright2019-zj, Tomesh2022-nu, Murali2019-my, Donkers2022-wt, Finzgar2022-aa, Mills2020-zh, Lubinski2023-zy, Lubinski2024-ci, Lubinski2023-mr, Chen2023-la, Benedetti2019-pp, Li2020-ry, Quetschlich2023-bg, Dong2021-gj, Martiel2021-vp, Van_der_Schoot2022-gv, Van_der_Schoot2023-vo, Cornelissen2021-yt, Georgopoulos2021-hh, Dong2022-ga}. 

Application-oriented quantum computer benchmarks are designed around quantum algorithms for computational problems~\cite{Proctor2025-cd}, such as factoring~\cite{Shor1994-zh} or Hamiltonian simulation~\cite{granet_analytical_2019,low_complexity_2023,motta_emerging_2022,low_optimal_2017,cao_quantum_2019,mcardle_quantum_2020,somma_quantum_2003,whitfield_simulation_2011,childs_toward_2018,lloyd_universal_1996}. A variety of application-oriented benchmarking suites have been developed~\cite{Chen2022-dm, Linke2017-mr, Wright2019-zj, Tomesh2022-nu, Murali2019-my, Donkers2022-wt, Finzgar2022-aa, Mills2020-zh, Lubinski2023-zy, Lubinski2024-ci, Lubinski2023-mr, Chen2023-la, Benedetti2019-pp, Li2020-ry, Quetschlich2023-bg, Dong2021-gj, Martiel2021-vp, Van_der_Schoot2022-gv, Van_der_Schoot2023-vo, Cornelissen2021-yt, Georgopoulos2021-hh, Dong2022-ga}, such as the benchmarking suite of the Quantum Economic Development Consortium (QED-C)~\cite{lubinski_application-oriented_2023,lubinski_optimization_2024,lubinski_quantum_2024,chatterjee_comprehensive_2025,niu_practical_2025} and SupermarQ~\cite{Tomesh2022-nu}. Unlike benchmarks for individual one- and two-qubit quantum gates (such as one- and two-qubit RB~\cite{Magesan2012-dz}), algorithmic benchmarks are sensitive to errors that only emerge in many-qubit circuits, like many-qubit crosstalk~\cite{Sarovar2020-pz, Gambetta2012-zd, Proctor2019-gf, Hines2023-tz, McKay2023-bx, Proctor2022-yl, Hines2023-vq}. Furthermore, unlike benchmarks that use many-qubit random circuits~\cite{Cross2019-ku, Magesan2011-hc, Proctor2022-yl, Hines2023-tz}, algorithm-oriented benchmarks can quantify the size and impact of errors in circuits that contain the same structures as algorithms. However, many existing application-based benchmarks are not scalable or have technical flaws caused by design decisions taken to enable scaling~\cite{Proctor2025-cd}.

Many application-oriented (and other) benchmarks rely on classical computations that scale exponentially in the number of qubits ($n$), e.g., because the benchmark requires classically computing the error-free output distribution of the circuit. Some application-oriented benchmarks use bespoke modifications to the application circuit to make the benchmark efficient~\cite{lubinski_application-oriented_2023, Proctor2025-cd}. However, this can lead to benchmarks that do not accurately predict performance on the original quantum circuits of interest, e.g., because the altered circuit has different sensitivity to errors \cite{Proctor2025-cd}. Furthermore, even a robustly-designed algorithmic benchmarking suite must choose which quantum algorithms and circuits to represent---they cannot include every interesting quantum circuit or algorithm. There is therefore a need for a tool that can create efficient and reliable benchmarks from user-chosen algorithms or circuits.

In the last few years, a variety of techniques have been developed that could enable such a user-friendly benchmarking tool, including \textit{mirror circuit fidelity estimation} (MCFE)~\cite{Proctor2021-wt, Proctor2022-zs, Proctor2022-yl}, \textit{full-stack MCFE}~\cite{hines_scalable_2024}, \textit{Cliffordization}~\cite{merkel_when_2025}, \textit{accreditation}~\cite{ferracin_reducing_2018,ferracin_accrediting_2019,Ferracin2021-vh}, and \textit{subcircuit volumetric benchmarking} (SVB)~\cite{seritan_benchmarking_2025}. These techniques are general-purpose tools for efficiently estimating a quantum computer's performance on some circuit, with complementary properties, e.g., different assumptions about a system's noise and different costs to implement. They enable almost any algorithm to be turned into a principled benchmark that can be both applied to contemporary systems and scaled to thousands or even millions of qubits. However, deploying these techniques to create new benchmarks has so far required detailed understanding of these methods and bespoke code, because no user-friendly implementation has been publicly available.

In this paper, we introduce and demonstrate software that implements many of these existing methods, enabling the creation of scalable benchmarks from quantum algorithms. Our software---called \oursoftware, for \textit{\textbf{sca}lable and \textbf{r}obust quantum \textbf{a}lgorithmic \textbf{b}enchmark generator}---is designed to enable the easy creation and implementation of benchmarks from \textit{user-chosen} quantum algorithms or circuits. \oursoftware's use does not require detailed knowledge of the benchmarking techniques it implements, e.g., an understanding of the theory of MCFE. It can therefore be used by quantum algorithms experts or other potential quantum computer users to design reliable and scalable quantum computer benchmarks, or to simply test if a particular quantum computer can successfully run an algorithm or circuit of interest to them. To demonstrate \oursoftware, we use it to create benchmarks that interrogate hardware and algorithmic tradeoffs in Hamiltonian simulation, that quantify the in-situ efficacy of approximate circuit compilation for algorithmic circuits, and that measure progress towards executing a (potentially very large) circuit of interest. 

\begin{figure*}[ht!]
    \centering
    \includegraphics[width=\linewidth]{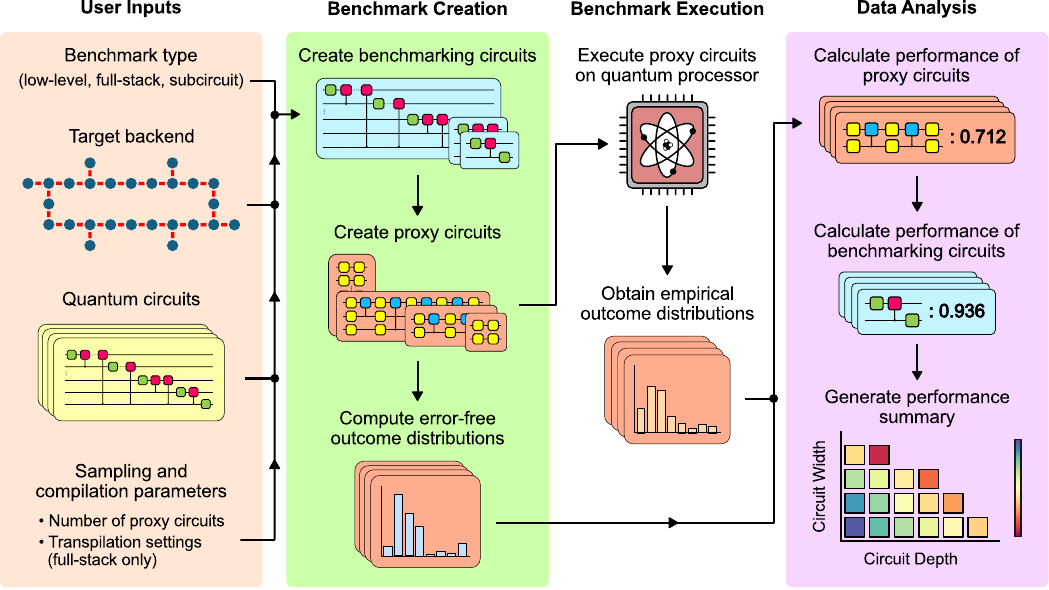}
    \caption{\begin{small}\textbf{Scalable benchmarking using \oursoftware.} A schematic of \oursoftware, which is software for creating efficient benchmarks from interesting quantum circuits on any number of qubits. \oursoftware~takes user-specified circuits, along with other options which we describe further in the main text, and creates an efficient and robust benchmark from those circuits. The benchmark consists of a set of benchmarking circuits $B$ whose performance is to be estimated, coupled with a set of proxy circuits $P$ that enable the efficient performance estimation of each circuit in $B$. Each proxy circuit is a mirror circuit, which enables the efficient classical computation of its error-free outcome distribution. The proxy circuits are then executed on the target quantum processor to obtain an empirical outcome distribution for each circuit in $P$. The empirical and error-free outcome distributions for the proxy circuits are then passed into the \oursoftware~data analysis function, which first calculates the performance of each proxy circuit and then uses the performance of the proxy circuits to calculate the performance of the benchmarking circuits. \oursoftware~also enables the creation of performance summaries including volumetric benchmarking and capability region plots~\cite{Proctor2021-wt}.\end{small}}
    \label{fig:flowchart}
\end{figure*}

\section{Preliminaries}
\subsection{Fidelity}
Our software is designed to create efficient benchmarks that measure process fidelity ($F$). We now review the definitions of $F$ and, for the purposes of comparison with existing benchmarks, classical fidelity. Process fidelity quantifies the accuracy with which a quantum process is implemented by a quantum processor~\cite{Hashim2024-om}, making it a well-motivated success metric, and is defined between two $n$-qubit superoperators $\mathcal{U}$ and $\Lambda$. Herein, $\mathcal{U}$ is the unitary evolution that some $n$-qubit circuit $c$ ideally implements (i.e., $\mathcal{U}[\rho] = U \rho U^{\dagger}$ where $U \in U(2^n)$) and $\Lambda$ is the superoperator corresponding to a noisy implementation of $c$. The process fidelity between such
$\mathcal{U}$ and $\Lambda$ is \cite{Hashim2024-om}
\begin{equation} \label{eq:def-process-fidelity}
    F(\mathcal{U}, \Lambda) = \frac{1}{4^n}\tr \left( \mathcal{U}^\dagger \Lambda \right).
\end{equation}
Process fidelity is widely used to quantify how well a circuit $c$ has been implemented~\cite{Hashim2024-om, Proctor2021-wt, Proctor2022-zs, hines_scalable_2024, merkel_when_2025, seritan_benchmarking_2025} and to quantify the error in individual gates \cite{Hashim2024-om}.

Instead of process fidelity, many existing benchmarks compute success metrics that compare a circuit's observed classical outcome distribution ($\tilde{p}$) to its ideal, error-free outcome distribution ($p$). A common metric of this sort is the classical fidelity between $\tilde{p}$ and $p$, given by
\begin{equation} \label{eq:def-classical-fidelity}
    F_c(p, \tilde{p}) = \left( \sum_{x } \sqrt{p(x) \tilde{p}(x)} \right)^2,
\end{equation}
where $x \in \{0,1\}^n$, $p(x)$ is the probability of obtaining the bitstring $x$ when the circuit is executed without error, and $\tilde{p}(x)$ is the probability (or, in practice, an estimate of the probability from observed frequencies) with which the bitstring $x$ is obtained in a noisy circuit execution. In the case of the QED-C's benchmarking suite, this fidelity is rescaled to the \textit{normalized} classical fidelity \cite{lubinski_application-oriented_2023}:
\begin{equation} \label{eq:def-normalized-fid}
    \bar{F_c}(p, \tilde{p}) = \frac{F_c\left(\tilde{p}, p\right) -  \frac{1}{2^n}  \sum_{x}\sqrt{p\left(x \right)}}{1 - \frac{1}{2^n} \sum_{x}  \sqrt{p\left(x \right)}}.
\end{equation}
Note that computing $p$ is generally exponentially costly, and that the normalized classical fidelity $\bar{F_c}$ is ill-behaved if $p$ is close to the uniform distribution.

\subsection{Mirror circuit fidelity estimation}
In our benchmarks, we estimate the process fidelity using MCFE~\cite{Proctor2022-zs}. MCFE estimates the process fidelity for $c$ by running three different kinds of mirror circuits, which we refer to as $M_1$, $M_2$ and $M_3$. The $M_1$ circuits consist of $c$ followed by a randomized compilation of its layer-by-layer inverse, surrounded by randomized single-qubit gates that implement randomized state preparation and measurement (SPAM). The $M_1$ circuits would enable estimating the process fidelity of $c$ if the inverse and SPAM were error-free, and the $M_2$ and $M_3$ circuits enable estimating the fidelity of the inverse circuit and the SPAM, so that $c$'s process fidelity can be isolated. The $M_2$ circuits do this by running a randomized compilation of $c$ and $c$'s layer-by-layer inverse, and the $M_3$ circuits are simple randomized SPAM circuits.

All three kinds of circuit contain randomized gates, and so when running MCFE is it necessary to decide how many circuits $|M_1|$, $|M_2|$, and $|M_3|$ of each kind to create. Increasing $|M_1|$, $|M_2|$ and $|M_3|$ increases the precision of the fidelity estimate, so we report $|M_1|$, $|M_2|$ and $|M_3|$ for each benchmark we create. \oursoftware~computes uncertainties on its process fidelity estimates using a non-parametric bootstrap, enabling the user to discern if an increase in $|M_1|$, $|M_2|$ and $|M_3|$ is necessary. We refer the reader to Refs.~\cite{Proctor2021-wt, Proctor2022-zs, Proctor2022-yl} for a more thorough presentation of MCFE.

\section{\texttt{scarab}: An efficient benchmark generator}
In this section we introduce \oursoftware, which is summarized in Fig.~\ref{fig:flowchart}. \oursoftware~is a high-level component within \texttt{pyGSTi} \cite{Nielsen2020-rd}, an open-source Python package that implements a broad suite of quantum characterization, verification, and validation (QCVV) \cite{Blume-Kohout2025-mp, Hashim2024-om} methods. \oursoftware~is a benchmark \textit{generator}---it creates a benchmark from user-given inputs. The inputs to \oursoftware~are:
\begin{itemize}
    \item \textit{The benchmark type.---}The kind of benchmark to be created. There are three possible kinds of benchmark: \textit{low-level}, \textit{full-stack}, and \textit{subcircuit benchmarks}. The different kinds of benchmark measure different aspects of performance, and they are detailed below.
    \item \textit{A quantum computer specification.---}A quantum computing system (a.k.a., ``backend''), which can be real or hypothetical, for which the benchmark is to be created.
    \item \textit{Quantum circuits.---}The quantum circuits $C$ from which to create the benchmark, specified as \texttt{qiskit} circuits. These are circuits that contains gates that are (ideally) unitary. The required level of abstraction of the circuits, i.e., whether they are high-level circuits needing compilation or low-level circuits for that system, depends on the benchmark type.
    \item \textit{Sampling and compilation parameters.---}Parameters that specify the number of circuits used in the fidelity estimation routines that \oursoftware~uses and (for full-stack benchmarks) compiler parameters. All such parameters have reasonable default options.
\end{itemize}

The \oursoftware~API (summarized in Fig.~\ref{fig:flowchart}) has two parts: circuit creation and data analysis. The circuit creation part takes the above inputs and creates a set of \textit{benchmarking circuits} ($B$) and, from them, creates a set of \textit{proxy circuits} ($P$), to execute. \oursoftware~efficiently estimates the process fidelity of each benchmarking circuit using data from the proxy circuits. The simplest case is $B = C$, i.e., the benchmark created is designed to directly measure the performance of the tested hardware on the input circuits $C$, which is the case in \oursoftware's low-level benchmarks. We discuss the form of $B$ for full-stack and subcircuit benchmarks below.

The benchmarking circuits $B$ are not the circuits that are output for execution by \oursoftware~because simply executing the circuits in $B$ will not enable the efficient estimation of these circuits' fidelities. Instead, \oursoftware~transforms $B$ into the set of proxy circuits $P$ that are to be executed, specified in \texttt{pyGSTi}'s circuit format (and which can be converted to \texttt{OpenQASM} or \texttt{qiskit} circuits using \texttt{pyGSTi}'s conversion functions). These proxy circuits are designed so that the performance of each circuit in $P$ can be \textit{efficiently} estimated and the process fidelity of each benchmarking circuit can be estimated from the performance of the proxy circuits. Each circuit in $P$ is created by applying MCFE~\cite{Proctor2022-zs} to the circuits in $B$, and therefore each circuit in $P$ is a \textit{mirror circuit} ~\cite{Proctor2021-wt, Proctor2022-zs}. We plan to add the option to instead use Cliffordization~\cite{merkel_when_2025}, an alternative fidelity estimation approach that uses Clifford proxy circuits instead of mirror circuits, in a future version of \oursoftware.

The proxy circuits output by $P$ must be executed by the user on their target quantum computing system (with no further compilation applied), recording the bit string observed in each execution of each circuit in $P$. That data is then processed by the data analysis function of \oursoftware, producing an estimate for the process fidelity of each circuit in $B$ (stored in a \texttt{pandas} \texttt{DataFrame} along with relevant metadata for the circuits) that can be easily manipulated by the user for custom analyses. \oursoftware~also contains a variety of built-in routines for results analysis and presentation, including \textit{volumetric benchmarking} and \textit{capability region} plots~\cite{Proctor2021-wt}.

We now explain the three kinds of benchmark that \oursoftware~can create:

\begin{itemize}
\item \textit{Low-level benchmarks.---}These benchmarks quantify the amount of noise in a set of input \textit{low-level} circuits $C$. They do so by efficiently measuring the process fidelity of each circuit in $C$. The input circuits $C$ must all already be compiled for the target system. For this benchmark type, \oursoftware's benchmarking circuits $B$ are simply the input circuits ($B=C$). The input circuits could be created by passing some high-level algorithmic circuits through a target system's built-in compilation algorithms, or the user may directly define low-level circuits of interest.

\item \textit{Full-stack benchmarks.---}These benchmarks efficiently measure the joint performance of a system's compiler and qubits. They do so by taking a set of input \textit{high-level} circuits $C$, compiling them to the target system to create the benchmarking circuits $B$, and then measuring the process fidelity of those compiled (i.e., transpiled and routed) circuits to the intended, exact unitaries defined by $C$. This kind of benchmark can test both exact and approximate compilation algorithms, as demonstrated in Section~\ref{sec:fullstack-transpiler-opt}.  \oursoftware's full-stack benchmarks currently use the \texttt{qiskit} transpiler with user-chosen transpiler flags, but future enhancements may enable flexibility in the compilation software used.

\item \textit{Subcircuit benchmarks.---}These benchmarks efficiently measure the performance of varied-shape (i.e., varied in width and depth) subcircuits sampled from the input circuits, enabling benchmarking of contemporary systems with utility-scale circuits \cite{seritan_benchmarking_2025}. For this type of benchmark, the input circuits $C$ must be already compiled for the target system, and the benchmarking circuits $B$ are varied-shape subcircuits from each circuit in $C$. Tested circuit shapes can be specified by the user.
\end{itemize}

\section{Runtime scaling} \label{sec:runtime-scaling}
\oursoftware~is designed to enable the creation of scalable benchmarks, and we demonstrate this scalability by measuring the runtime of \oursoftware. In particular, we measured the \textit{classical processing time} ($t_c$), which we define to be the time taken to generate the benchmark's proxy circuits $P$, compute any information that is needed to process the data from those circuits (which, in the case of \oursoftware's benchmarks, is a target bit string for each proxy circuit), and perform the data analysis to determine the performance of both the proxy circuits $P$ and the benchmarking circuits $B$. The time $t_c$ therefore encompasses the complete classical computational cost of creating and analyzing the results of \oursoftware's benchmarks.

\begin{figure}[t!]
    \centering
    \includegraphics[width=\linewidth]{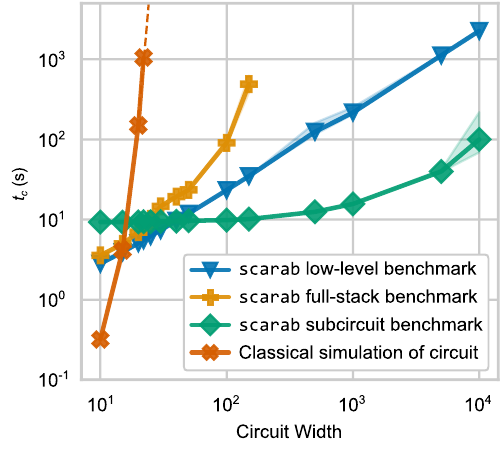}
    \caption{\begin{small}\textbf{Classical processing time for \oursoftware.} The time ($t_c$) taken by the classical processing in \oursoftware, consisting of the time to turn a circuit input into \oursoftware~into the ``proxy circuits'' to be run, compute all the information about those circuits needed to analyze data from those circuits (e.g., error-free outcome distributions), and to calculate the performance of the proxy and benchmarking circuits using the \oursoftware~data analysis function; versus the circuit's width (number of qubits, $n$). We show the mean $t_c$ (markers), and the best and worst $t_c$ (shaded region) over different circuits, for the three kinds of benchmarks created by \oursoftware: low-level benchmarks (blue triangles), full-stack benchmarks (yellow pluses), and subcircuit benchmarks (green diamonds). We compare the scaling of $t_c$ for \oursoftware~to the time to classically simulate the same quantum circuits using \texttt{qiskit\_aer} (orange crosses) from which we created efficient benchmarks with \oursoftware. This classical simulation is a key step in many other benchmarks, and, unlike the scaling of $t_c$ for \oursoftware, the classical simulation scales exponentially (fit line) and is therefore impractical for many-qubit circuits.\end{small}}
    \label{fig:runtime-scaling}
\end{figure}

We measured $t_c$ for the three different kinds of benchmark that we can create with \oursoftware: low-level benchmarks, full-stack benchmarks, and subcircuit benchmarks. \oursoftware~creates a benchmark from a user-given circuit, and so to compute $t_c$ we must select a circuit or circuit family. We computed $t_c$ for $n$-qubit U3-CZ brickwork circuits~\cite{merkel_when_2025} of depth 128 with $n$ varied. We set $|M_1| = |M_2| = |M_3| = 10$. For each mirror circuit, we generate fake shot data (in order to compute the data processing time) from a uniform distribution over all $2^n$ bit strings with 1024 shots. Fig.~\ref{fig:runtime-scaling} shows the mean $t_c$ for each kind of benchmark versus $n$. For low-level and subcircuit benchmarks we measured $t_c$ up to $n=10000$. We observe that $t_c$ scales linearly for the low-level benchmarks (blue triangles, Fig.~\ref{fig:runtime-scaling})  and is still practical ($t_c \approx 5000\, \textrm{s}$) even for $n=10000$. For the subcircuit benchmarks, we observe that $t_c$ scales sublinearly (green diamonds, Fig.~\ref{fig:runtime-scaling}). This sublinear scaling arises from the fact that, in this test, we are creating benchmarks whose proxy circuits are subcircuits of the input $n$-qubit circuit with an $n$-independent shape (we create subcircuits of shapes $\{(w_i, d_i) \} = \{2,4,6\} \times \{2,4,8\}$).

To test the runtime of \oursoftware's full-stack benchmarks, we chose IBM Fez as the target quantum backend and used the \texttt{qiskit} transpiler with \texttt{optimization\_level} set to 3. The qubit count of IBM Fez limits these numerical experiments to 150 qubits. The time $t_c$ for the full-stack benchmark (orange squares, Fig.~\ref{fig:runtime-scaling}) is still practical even at $n=150$ ($t_c \approx 800\, \textrm{s}$). We also report the time taken to classically simulate the U3-CZ brickwork circuits we used as input to \oursoftware~(orange crosses, Fig.~\ref{fig:runtime-scaling}). That simulation, implemented using \texttt{qiskit} and which is required by many other benchmarks, has an exponentially growing cost. In contrast, the classical processing cost of \oursoftware~benchmarks makes them practical even with utility-scale circuits.

\begin{figure}[th!]
    \centering
    \includegraphics[width=\linewidth]{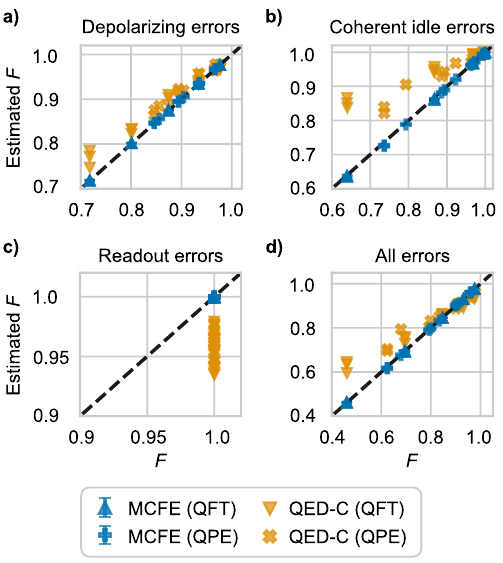}
    \caption{\begin{small}\textbf{Estimating process fidelity with \oursoftware.} Simulations demonstrating that benchmarks created with \oursoftware~reliably estimate a circuit's process fidelity $F$ in the presence of complex device errors. We show the $F$ estimated using \oursoftware~benchmarks (blue markers) versus the
    true $F$ for two classes of circuit (QFT and QPE circuits) under different noise models. Error bars are one standard deviation and calculated using a non-parametric bootstrap.
    To demonstrate the difference between process fidelity $F$ and the normalized classical fidelity $\bar{F_c}$ estimated by other benchmarks, we also plot $\bar{F_c}$ versus $F$ (orange markers) for these circuits. We observed that \oursoftware~benchmarks accurately estimate the process fidelity in the presence of (a) depolarizing errors on gates, (b) coherent errors on gates, (c) readout errors, and (d) all three kinds of errors. The normalized classical fidelity can be significantly larger or smaller than the process fidelity, depending on the details of the noise model.\end{small}}
    \label{fig:classical-mcfe-pf}
\end{figure}

\section{Demonstrating reliable fidelity estimation}
\label{sec:exact-vs-meas-pf}
We now demonstrate that \oursoftware~reliably estimates the process fidelity $F$, and we explain how this differs from metrics based on classical fidelity used in other benchmarks. To demonstrate that \oursoftware's benchmarks reliably estimate process fidelity, we simulated benchmarks created using \oursoftware~and compared the estimated process fidelities to the true values of $F$. For these demonstrations, the input circuits to \oursoftware~are circuits from the QED-C's Application-Oriented Benchmarking suite~\cite{lubinski_application-oriented_2023} (see the appendices for a summary of that suite). This is therefore also a demonstration of using \oursoftware~to convert an existing benchmark into a \textit{scalable} benchmark with a well-motivated success metric: process fidelity.

In these simulations, we used $n$-qubit circuits with $n=3$ to $n=6$ from the quantum phase estimation (QPE) and the quantum Fourier transform (QFT) benchmarks of the QED-C's suite. We set $|M_1| = |M_2| = |M_3| = 1000$, and we use \texttt{qiskit\_aer} to simulate the noisy circuits. The circuits are compiled to, and the noise model is defined on, the native gate set for many IBM Q systems: the $\{X, SX, RZ, CZ\}$ gate set. Fig.~\ref{fig:classical-mcfe-pf} (blue markers) compares the true process fidelity to the estimate from \oursoftware. We make these comparisons for four different noise models (detailed further below), corresponding to the four panels in the figure. For all noise models, the \oursoftware~process fidelity estimate is in close agreement with the true process fidelity (error bars are 1 standard deviation, and in most cases they are smaller than the data markers). This demonstrates using \oursoftware~to convert existing unscalable benchmarks into scalable benchmarks, and that \oursoftware~creates benchmarks that robustly estimate process fidelity.

Many existing benchmarks---including many of the QED-C benchmarks---compute success metrics based on the classical fidelity. Classical fidelity and process fidelity measure different aspects of performance.  Unlike classical fidelity, process fidelity is sensitive to errors that will affect any input state and final measurement, whereas the classical fidelity is computed for a particular input state and measurement. The process fidelity is therefore more appropriate in cases where a circuit will be used as a subroutine---like QPE and the QFT---and is also arguably more relevant when the circuits being used to create benchmarks are proxies for larger utility-scale circuits (e.g., a small QFT being a proxy for a large QFT used in a useful algorithm). By construction, process fidelity does not quantify SPAM errors, whereas classical fidelity includes a contribution from those errors. We note, however, that \oursoftware's benchmarks do enable estimation of SPAM error (and separating it from circuit error), although \oursoftware's analysis pipeline does not currently report estimates of SPAM errors.

\begin{figure*}[!ht]
    \centering
    \includegraphics[width=\linewidth]{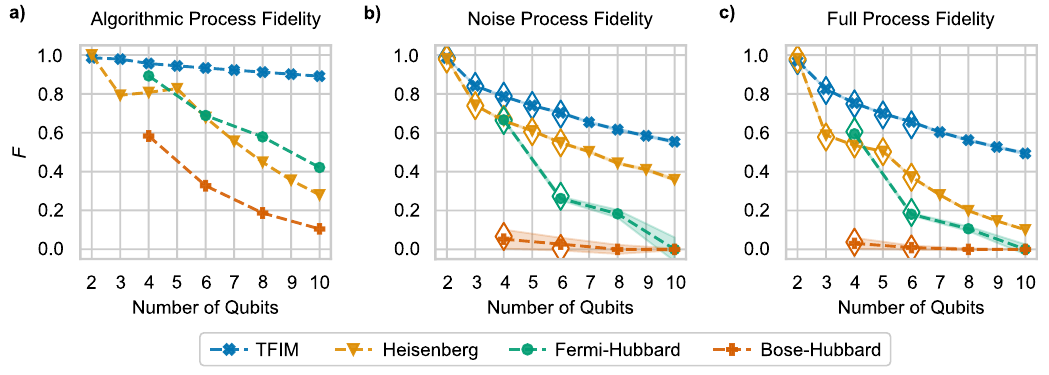}
    \caption{\begin{small} \textbf{Estimating the impact of noise and algorithm approximation with \oursoftware's efficient low-level benchmarks.} Using \oursoftware, we created low-level benchmarks from first-order Trotter circuits with four different $n$-qubit Hamiltonians---TFIM, Heisenberg, Fermi-Hubbard, and Bose Hubbard Hamiltonians---from \texttt{HamLib}. (a) The algorithmic process fidelity, i.e., the fidelity between the Trotter circuit's (noise-free) unitary and the ideal unitary evolution for that Hamiltonian, versus $n$. (b) The noise process fidelity estimated by the \oursoftware~benchmarks (solid markers), which is the process fidelity between the noisy Trotter circuit and the noise-free unitary that circuit implements, and its exact value (open diamonds) up to $n=6$. (c) The estimated full process fidelity (solid markers)---i.e., the process fidelity between the ideal unitary evolution and the noisy implementation of the Trotter evolution, approximated as the product of the measured noise process fidelity and the computed algorithmic process fidelity---and its exact value (open diamonds) up to $n=6$. For both the noise and full process fidelities, we observe close agreement between the \oursoftware~estimate and the true values. 
    Shaded regions around the \oursoftware~estimates for the process fidelities are 1 standard deviation calculated from a non-parametric bootstrap. \end{small}} \label{fig:hsqs}
\end{figure*}

We demonstrate the difference between classical fidelity and process fidelity in  Fig.~\ref{fig:classical-mcfe-pf}. We show how normalized classical fidelity (orange markers) deviates from process fidelity under different noise models. Fig.~\ref{fig:classical-mcfe-pf}(a) compares process fidelity and normalized classical fidelity for a noise model with depolarizing noise of strength $\lambda_{1Q} = 0.0005$ and $\lambda_{2Q} = 0.005$ on all one- and two-qubit gates, respectively. The normalized classical fidelity overestimates the process fidelity by up to 0.07 or 10\%. Fig.~\ref{fig:classical-mcfe-pf}(b) shows the comparison for a noise model with only readout error, with a rate of $\epsilon=0.01$. The process fidelity is unity in this case, but $\bar{F_c}$ is sensitive to readout error and therefore underestimates the process fidelity by up to 0.1 or 10\%. Fig.~\ref{fig:classical-mcfe-pf}(c) shows the comparison for a noise model where every idle gate experiences a $\theta_{\textrm{idle}} = 0.005$ radian $Z$ rotation. These coherent idle errors create phase errors before computational basis measurements, which are not captured by classical fidelity. As a result, the classical fidelity overestimates the process fidelity by up to 0.25, or 40\%, which is greater than in the case of depolarizing noise. Fig.~\ref{fig:classical-mcfe-pf}(d) shows the combined effect of all three error sources: depolarizing noise, readout error, and idle rotation. For smaller circuits, the effect of the depolarizing noise and idle rotation is smaller and the readout error is dominant, causing the classical fidelity to underestimate the process fidelity. For larger circuits, the depolarizing errors and idle rotations contribute more to the total error in the circuit, which leads the classical fidelity to overestimate the process fidelity.

\begin{figure*}[ht!]
    \centering
\includegraphics[width=\linewidth]{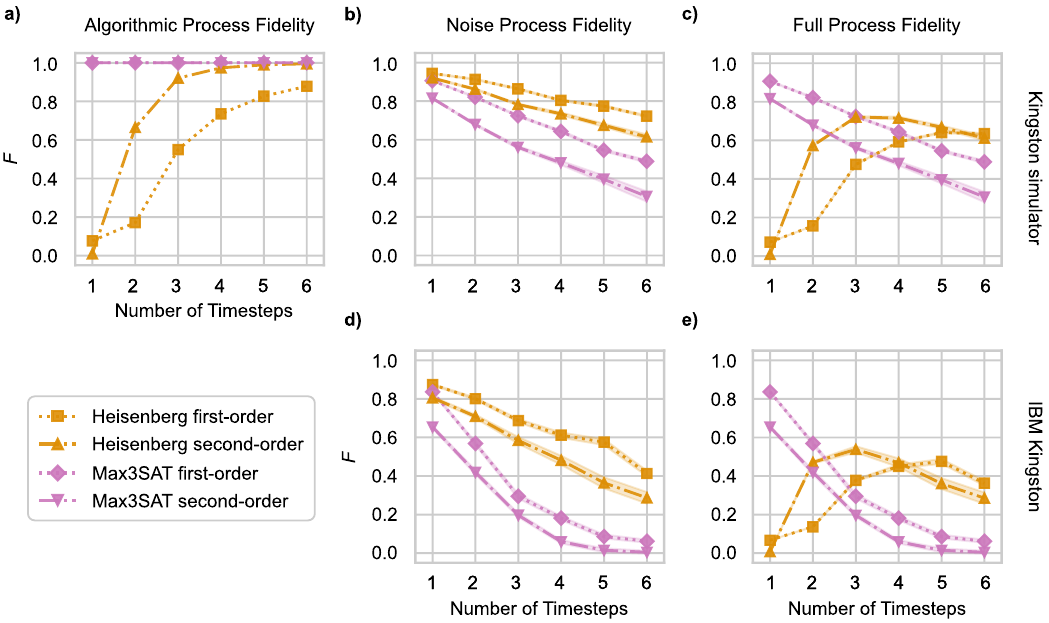}
    \caption{\begin{small}\textbf{Quantifying noise and algorithmic tradeoffs using \oursoftware~benchmarks.} Using \oursoftware, we created low-level benchmarks from first-order and second-order Trotter circuits for two 5-qubit Hamiltonians (Heisenberg and Max3SAT) with varying number of time steps. These benchmarks were created for IBM Kingston, and both run on IBM Fez and IBM's simulator of IBM Kingston. (a) The algorithmic process fidelity versus the number of time steps for each Hamiltonian and both first- and second-order Trotter circuits. We used \oursoftware~to estimate the noise process fidelity with (b) the simulator of IBM Kingston and (d) IBM Kingston. In all cases, we observe that the noise fidelity decreases as the number of time steps increases, due to increasing depth of the circuit. To quantify the optimal trade-off between noise and algorithmic fidelity, we estimated the full process fidelity for (c) the simulation of IBM Kingston and (d) IBM Kingston. For the Heisenberg Hamiltonian, we find that the optimal algorithm parameters are a second-order Trotter circuit with 3 time steps, in both the simulation and the experiment. Shaded regions around the \oursoftware~estimates for the process fidelities are 1 standard deviation calculated from a non-parametric bootstrap.\end{small}}
    \label{fig:hsts}
\end{figure*}

\begin{figure*}[ht!]
    \centering
    \includegraphics[width=\linewidth]{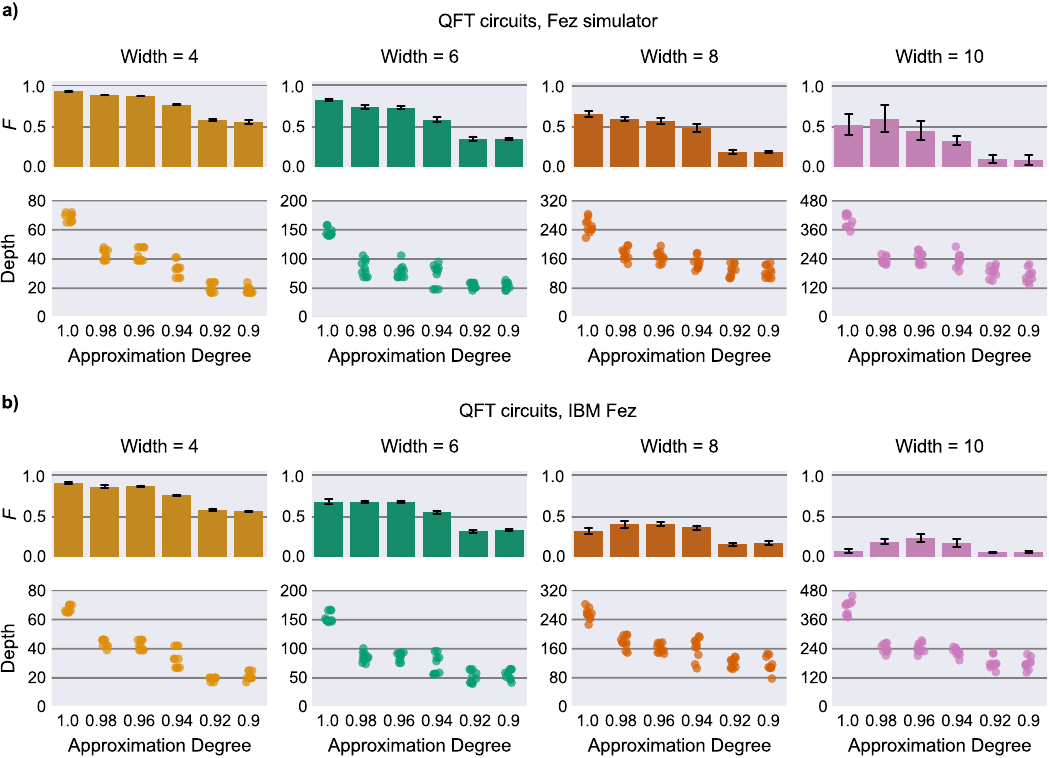}
    \caption{\begin{small}\textbf{Testing approximating compilation algorithms using \oursoftware's full-stack benchmarks for the QFT.} The efficacy of approximate circuit compilation for QFT circuits, for IBM Fez and a simulation of IBM Fez that uses IBM's noise model for this system. We show estimates of the process fidelities of compiled QFT circuits, to the ideal, approximation-free unitaries, for $n=2$ up to $n=10$ qubits and with varying approximation degree in the compilation algorithm (error bars are one standard deviation). We also show the depths of the compiled circuits for each approximation degree. Error bars represent a 95\% confidence interval calculated via a nonparametric bootstrap.\end{small}}
    \label{fig:fullstack-qft}
\end{figure*}

\begin{figure*}[ht!]
    \centering
    \includegraphics[width=\linewidth]{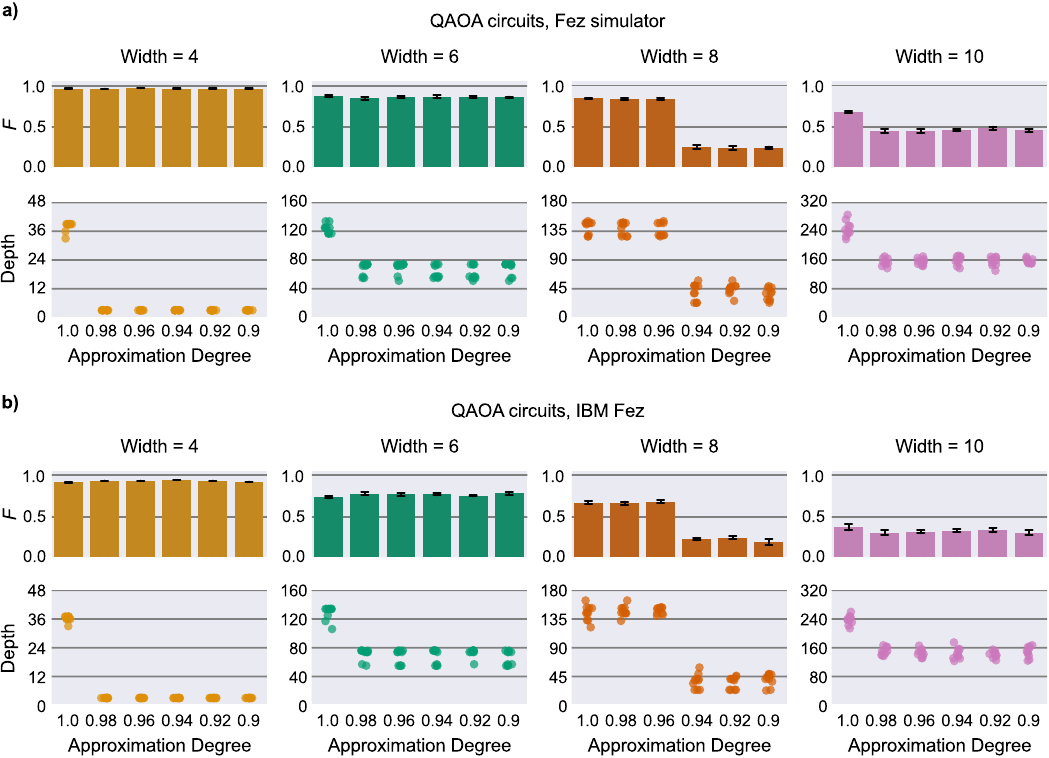}
    \caption{\begin{small}\textbf{Testing approximating compilation algorithms using \oursoftware's full-stack benchmarks for QAOA.} The efficacy of approximate circuit compilation for QAOA circuits, for IBM Fez and a simulation of IBM Fez that uses IBM's noise model for this system. We show estimates of the process fidelities of compiled QAOA circuits, to the ideal, approximation-free unitaries, for $n=2$ up to $n=10$ qubits and with varying approximation degree in the compilation algorithm (error bars are one standard deviation). We also show the depths of the compiled circuits for each approximation degree. Error bars represent a 95\% confidence interval calculated via a nonparametric bootstrap.\end{small}}
    \label{fig:fullstack-qaoa}
\end{figure*}

\section{Scalable Hamiltonian simulation benchmarks}
In this and the following two sections, we showcase three applications for \oursoftware, using each kind of benchmark that \oursoftware~can create. We show how \oursoftware's benchmarks can be used to efficiently interrogate the performance of quantum computers and how hardware noise and algorithmic error combine to impact a quantum algorithm. In each case, we demonstrate \oursoftware~using both simulations and IBM Q experiments.

\subsection{Hamiltonian simulation} \label{sec:ham-sim}
In our first application of \oursoftware, we show how it can be used to create low-level benchmarks for Hamiltonian simulation circuits. These benchmarks are designed to enable exploration of the hardware and algorithmic accuracy trade-offs for Hamiltonian simulation, where more accurate algorithms can be used at the cost of more gates---and therefore more noise.

In Hamiltonian simulation algorithms, a core subroutine is the approximate implementation of a unitary $U$ generated by applying a Hamiltonian $H$ for some time $t$, given by 
\begin{equation}
U = \exp(-iHt).
\end{equation}
Hamiltonian simulation algorithms do not typically apply $U$ exactly, but instead some unitary $\widetilde{U}$ that approximates $U$, i.e., $\widetilde{U} \approx U$. The approach to approximating $U$ that we consider here is Trotterization. If $H = \sum_{j=1}^k H_j$, then the first-order Trotterization is~\cite{suzuki_generalized_1976}
\begin{equation} \label{eq:def-first-order-trotter}
    \widetilde{U} = \left(\prod_{j=1}^k\exp(-i H_j t/m)\right)^m.
\end{equation}
The magnitude of the error in this approximation is given by
\begin{equation} \label{eq:first-order-trotter-error}
    U = \widetilde{U} + O\left(\frac{t^2}{m}\right).
\end{equation}
By increasing the number of Trotter steps ($m$), the discrepancy between $U$ and $\widetilde{U}$ can be decreased, but at the cost of deeper circuits.
In addition to the first-order Trotterization, we will also consider second-order Trotterization~\cite{suzuki_generalized_1976,beig_finding_2005}, where
\begin{equation} \label{eq:def-second-order-trotter}
    \widetilde{U} = \left[\left(\prod_{j=1}^k \exp\left(\frac{-i H_j t}{2m}\right)\right)\left(\prod_{j=1}^k \exp\left(\frac{-i H_{k+1-j} t}{2m}\right)\right)\right]^m.
\end{equation}
The magnitude of the error in second-order Trotterization is given by
\begin{equation} \label{eq:second-order-trotter-error}
    U = \widetilde{U} + O\left(\frac{t^3}{m^2}\right).
\end{equation}

There are three superoperators that are important for quantifying the performance of a noisy implementation of a Trotterization circuit $c$:
\begin{itemize}
 \item The superoperator  $\mathcal{U}$ giving the ideal Hamiltonian evolution with the action $\mathcal{U}[\rho]=U\rho U^{\dagger}$.
 \item The superoperator $\widetilde{\mathcal{U}}$ for the approximation to $U$ implemented by an error-free (i.e., noiseless) execution of the Trotterized circuit $c$, with the action  $\widetilde{\mathcal{U}}[\rho] = \widetilde{U}\rho\widetilde{U}^{\dagger}$.
 \item The noisy Trotter evolution $\Upsilon$, which is the superoperator corresponding to the noisy implementation of the circuit $c$.
\end{itemize}
These superoperators enable us to define three process fidelities that summarize different aspects of the error in a Trotterization circuit:
\begin{itemize}
 \item The \textit{algorithmic process fidelity} $F(\mathcal{U}, \widetilde{\mathcal{U}})$, which captures the error due to the Trotter approximation.  
 \item The \textit{noise process fidelity} $F(\widetilde{\mathcal{U}}, \Upsilon)$, which captures the error in implementing the Trotterized circuit due to hardware noise.
 \item The \textit{full process fidelity} $F(\mathcal{U}, \Upsilon)$, which captures the combine impact of both Trotter approximation error and hardware noise.
\end{itemize}

\oursoftware~enables efficiently measuring the noise process fidelity, by creating a low-level benchmark from the Trotter circuit $c$, which we demonstrate below with simulations and experimental data. First, however, we consider how we can estimate the full process fidelity. Directly computing the full process fidelity requires access to $\mathcal{U}$, which is generally infeasible as it is an exponentially-large matrix (and we also do not have access to $\Upsilon$ without exponentially-expensive tomography~\cite{Hashim2024-om}). We therefore propose approximating the full process fidelity as a  product of the noise process fidelity and algorithmic process fidelity:
\begin{equation}
    F(\mathcal{U}, \Upsilon) \approx F(\mathcal{U}, \widetilde{\mathcal{U}}) \cdot F(\widetilde{\mathcal{U}}, \Upsilon).
    \label{eq:fpf-approx}
\end{equation}
Computing the right hand side of Eq.~\eqref{eq:fpf-approx} directly requires access to the $O(2^n)$ unitary matrices $U$ and $\widetilde{U}$, to compute $F(\mathcal{U}, \widetilde{\mathcal{U}})$, which is not scalable. However, Trotter error is well studied~\cite{wiebe_higher_2010,childs_theory_2021}, and we conjecture that it may be possible to efficiently estimate the algorithmic process fidelity even for large $n$. 

\subsection{Demonstration in simulations}
We demonstrate scalable Trotterization benchmarks created with \oursoftware~and explore the accuracy of Eq.~\eqref{eq:fpf-approx} using simulations (all these benchmarks use $|M_1| = |M_2| = |M_3| = 100$). We consider first-order Trotterized circuits for the transverse field Ising model (TFIM), Heisenberg, Bose-Hubbard, and Fermi-Hubbard Hamiltonians on $n=2$ to $n=10$ qubits, using Hamiltonians from \texttt{HamLib}~\cite{sawaya_hamlib_2024}. More information on these Hamiltonians and the specific choice of parameters for each can be found in the appendices. We used code from the QED-C HamLib benchmark~\cite{chatterjee_comprehensive_2025} in order to select specific Hamiltonians from the HamLib datasets and generate the first-order Trotter circuits. The circuits are compiled to, and the noise model is defined on, the $\{X, SX, RZ, CZ\}$ gate set. We simulated a noise model with depolarizing errors on one- and two-qubit gates with depolarizing parameters of  $\lambda_{1Q} = 0.0005$ and $\lambda_{2Q} = 0.005$, respectively, with coherent over-rotation errors on the $X$ and $SX$ gates of $\theta_{X} = \theta_{SX} = 0.01$ radians, and with $\epsilon=0.01$ readout error. 

In Fig~\ref{fig:hsqs}, we show the (a) algorithmic, (b) noise, and (c) full process fidelities for these Trotter circuits and this error model. Fig.~\ref{fig:hsqs}(b) shows the noise process fidelity estimated using \oursoftware~low-level benchmarks created from these circuits (solid markers) and the exact value of the noise process fidelity for $n \leq 6$ qubits (open diamonds). As in our in simulations of Fig.~\ref{fig:classical-mcfe-pf}, we observe close agreement between \oursoftware's estimate of the noise process fidelity and its true value. In Fig.~\ref{fig:hsqs}(c), we show estimates of the full process fidelity from \oursoftware's estimates of the noise process fidelities and the exact algorithmic process fidelities in Fig.~\ref{fig:hsqs}(a), using Eq.~\eqref{eq:fpf-approx}. We compare these estimates (solid markers) to the exact values of the full process fidelities (open diamonds), again seeing close agreement.

\subsection{Demonstration in experiment}

We used \oursoftware~to generate Hamiltonian simulation benchmarks that explore how to maximize the accuracy of the Hamiltonian simulation in-situ, in experiments on IBM Kingston. We explore how to balance the accuracy in the approximation of $U$ with the effects of hardware error (see Refs.~\cite{knee_optimal_2015,jones_optimising_2019,avtandilyan_optimal-order_2024} for similar investigations). We can quantify this trade-off by estimating the full process fidelity, which is sensitive to both Trotter error and hardware noise. We used 5-qubit Heisenberg and Max3SAT Hamiltonians and Trotterization at first or second order. We used between 1 and 10 time steps ($m$) for each Trotterization order. We used \oursoftware~to create low-level benchmarks ($|M_1| = |M_2| = 200, |M_3| = 10$) from these Trotter circuits, and ran them on IBM Kingston as well as a \texttt{qiskit\_aer} simulation of IBM Kingston.

Figure~\ref{fig:hsts} shows the results of these experiments and simulations. Figure~\ref{fig:hsts}(a) shows the algorithmic process fidelity, versus the number of time steps, for each Trotter order and each Hamiltonian. This was calculated exactly using simulations. For the Heisenberg Hamiltonian, we observe that increasing the number of time steps decreases the Trotter error, as expected. For the Max3SAT Hamiltonian, all terms commute and therefore $\widetilde{U} = U$ so the algorithmic process fidelity is unity. Fig.~\ref{fig:hsts}(b) and (d) show the noise process fidelity in simulation and experiment, respectively, estimated using \oursoftware.

Increasing the order or the number of time steps decreases the noise process fidelity, because the circuits increasing in depth. However, increasing the order of the Trotterization or the number of time steps increases the algorithmic process fidelity in most cases (see Fig.~\ref{fig:hsts}(a)) for the Heisenberg Hamiltonian, and so an decrease in noise process fidelity could be offset by a larger increase in algorithmic process fidelity.
To quantify this, we estimated the full process fidelity (using Eq.~\eqref{eq:fpf-approx}), with these estimates shown in Fig.~\ref{fig:hsts}(c) and (e). For Max3SAT, the full process fidelity is maximized with 1 time step and first-order Trotterization---because the algorithmic fidelity is unity for all cases. However, for the Heisenberg Hamiltonian, we find that the full process fidelity is maximized with second-order Trotterization with 3 time steps.

\section{Testing compilers with scalable full-stack benchmarks}
\label{sec:fullstack-transpiler-opt}

We now demonstrate using \oursoftware~to create full-stack benchmarks that enable efficient in-situ testing of compiler optimizations that aim to balance the impacts of noise and algorithmic approximation. Compilation translates a circuit into a system's native gate set and topology, and compilation algorithms are typically designed with the aim of finding a circuit that will execute with low error (i.e., high fidelity) on that system, e.g., by minimizing circuit depth or the number of two-qubit gates. A compilation of a circuit can either be exact or approximate. An \textit{exact} compilation of a circuit $c$ creates another circuit $c'$ that implements the same unitary as $c$, i.e., $U(c') = U(c)$ where $U(c)$ and $U(c')$ are the unitaries implemented by the circuits $c$ and $c'$, respectively. In contrast, \textit{approximate} compilation of a circuit $c$ creates another circuit $c'$ that only approximately implements the same unitary as $c$, i.e., $U(c') \approx U(c)$. Approximate compilation is a weaker condition, potentially allowing for compiling the input circuit into a low-level circuit that will execute with higher fidelity, e.g., because the circuit is much shallower. This enables trading off intrinsic error---the difference between $U(c')$ and $U(c)$---and errors due to hardware noise. For instance, a compiler could discard small controlled rotations, or small-angle (e.g., $\pi/256$) single-qubit rotations, because those gates are expected to cause more error than accrued by not implementing them at all. This tradeoff has been investigated, e.g., in~\cite{kalloor_application_2025}.

We explored the in-situ performance of \texttt{qiskit}'s approximate compilation algorithm using scalable full-stack benchmarks ($|M_1| = |M_2| = |M_3| = 10$) created with \oursoftware, designed for and executed on IBM Fez. We created full-stack benchmarks from two kinds of algorithmic circuits: QFT circuits and quantum approximate optimization algorithm (QAOA) circuits (obtained from the \texttt{qiskit} library). For the QAOA circuits, the cost operator we used corresponds to a random $\text{GNP}(n, 2 \ln(n)/n$) graph (we used the default \texttt{qiskit} mixing operator). We used one repetition of each operator and initialize the mixing angles uniformly between $0$ and $\pi$. For compilation, we used the \texttt{qiskit} transpiler with \texttt{optimization\_level} set to 3. We varied the \texttt{approximation\_degree} parameter from $0.9$ to $1.0$, with 1.0 corresponding to exact compilation. We executed these benchmarks on IBM Fez as well as a \texttt{qiskit\_aer} simulation of IBM Fez. Since \texttt{qiskit}'s transpilation is stochastic (i.e., the same compiled circuit is not always produced for a fixed input circuit) we tested 10 transpilations of each high-level circuit. 

Figures~\ref{fig:fullstack-qft} and~\ref{fig:fullstack-qaoa} show the results of these experiments and simulations. In most cases, the process fidelity is not improved by using approximate compilation, i.e., the process fidelity for \texttt{approximation\_degree} $= 1.0$ is the largest or within 1$\sigma$ of the largest estimated process fidelity. However, for the experimental results with the QFT on 8 and 10 qubits, the maximum process fidelity is achieved with approximate compilation. The $n$-qubit QFT contains controlled rotations with angles that decrease with increasing $n$, and so as $n$ increases it is possible to improve the noisy QFT's process fidelity by dropping these gates---which, because they are close to the identity gate, has a only small impact on the intrinsic error of the circuit---rather than implementing them imperfectly. These full-stack benchmarks enable quantify this effect in situ, with a system's actual noise processes.

\begin{figure*}[ht!]
    \centering
   \includegraphics[width=\linewidth]{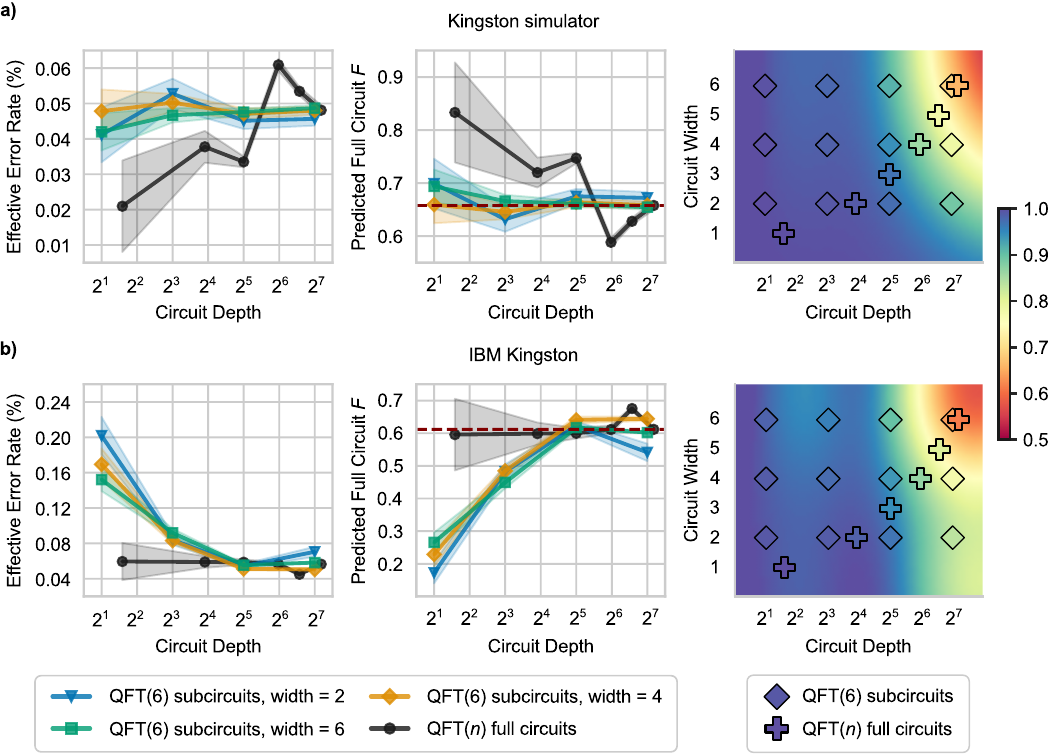}
    \caption{\begin{small} \textbf{Demonstrating scalable subcircuit benchmarking using \oursoftware.} \oursoftware~enables the creation of scalable benchmarks from a potentially-large input circuit $c$ in which the process fidelities of varied-shape subcircuits of $c$ are estimated. Here we demonstrate this in simulation (upper row) and experiment (lower row) for a small example: a 6-qubit QFT circuit, QFT(6). \oursoftware~also enables scalable benchmarks in which the input size of an algorithm is varied, and we also demonstrated this by varying the input size $n$ of the QFT, QFT($n$). The rightmost plots show the measured process fidelities of subcircuits of QFT(6) of each of 12 different shapes (diamonds) as well as the measured process fidelities of the QFT($n$) circuits (pluses). The heatmap is the predictions for process fidelity obtained by applying Gaussian process regression to the subcircuit data. The leftmost and central plots show effective error rates (see main text) and their predictions for the QFT(6)'s process fidelity, obtained from the subcircuits of each shape (colored points) and the QFT($n$) circuits (black lines). Error bars in the EER  and predicted full circuit process fidelity plots are one standard deviation calculated from a non-parametric bootstrapped distribution.\end{small}} 
    \label{fig:subcirc-pred}
\end{figure*}

\section{Predicting algorithm performance with subcircuit benchmarks}
\label{sec:subcirc-performance-pred}
In this section we demonstrate \oursoftware's subcircuit benchmarks and \oursoftware's ability to turn a algorithmic circuit directly into an efficient benchmark. Many interesting circuits are so large that they cannot be implemented with significantly non-zero fidelity on contemporary systems. For example, many of the most promising algorithms for quantum computers appear to require very deep circuits on thousands of qubits~\cite{Proctor2025-cd}. One approach to creating benchmarks that can be run on contemporary systems but that are built from potentially very large circuits is to run subcircuits ``snipped'' out of the given circuit(s)~\cite{seritan_benchmarking_2025}. This method is implemented by \oursoftware's subcircuit benchmarks. Performance on those circuits can potentially enable the prediction of performance on the input (potentially very large) circuit, and provide a principled approach to measuring progress towards implementing that large circuit with low error. Another approach, possible for circuits with tunable input size or number of qubits $n$, is to measure performance as $n$ is varied. This is also possible to do efficiently, using \oursoftware's low-level benchmarks.

To demonstrate both approaches to algorithmic benchmarking, we created a subcircuit benchmark from a 6-qubit QFT, and we also created low-level benchmarks from the $n$-qubit QFT with $n=2,3,4,5,6$. The subcircuit benchmarks used sampling parameters $|M_1| = |M_2| = 50, |M_3| = 100$ and the $n$-qubit QFT benchmarks used $|M_1| = |M_2| = |M_3| = 100$. We compiled the $n$-qubit QFT for IBM Kingston, used the compiled $6$-qubit circuit to create a subcircuit benchmark with \oursoftware, and used the compiled $n$-qubit QFTs to create low-level benchmarks with \oursoftware~directly from those circuits. We ran these benchmarks on IBM Kingston and also simulated them using the \texttt{qiskit\_aer} simulation of IBM Kingston. One of the parameters of the subcircuit benchmark type is the circuit shapes to use: we created subcircuits at shapes $(w,d) \in (2,4,6) \times (2^1, 2^3, 2^5, 2^7)$ and sampled 30 subcircuits of each shape. We chose the 6-qubit QFT because it is a small subroutine that we can directly run on contemporary systems, and this enable us to also estimate the process fidelity of the 6-qubit QFT (using \oursoftware) and compare its observed performance to extrapolations from subcircuit benchmarks, with each circuit shape, created from that 6-qubit QFT.

Figure~\ref{fig:subcirc-pred} shows the results of these simulations (panel (a)) and experiments (panels (b)). In the rightmost plot of Fig.~\ref{fig:subcirc-pred} (a) and (b) we show the estimated mean process fidelities of the subcircuits of each shape (diamonds), as well as the estimated process fidelities of the $n$-qubit QFT circuits (pluses), for the simulated and experimental data, respectively. These results are presented on a volumetric plot, i.e., shown on circuit depth $\times$ circuit width axes. We also show the predictions of a Gaussian process regression (GPR) model fit to the subcircuit data (heat map)~\cite{proctor_featuremetric_2025}. The measured process fidelities versus circuit shape, as well as the GPR-derived heat map, provide a visual summary of these system's performance on QFT circuits. Furthermore, this GPR model is one approach to predicting the performance of other circuits (e.g., the 6-qubit QFT) from observed process fidelities at a discrete set of circuit shapes \cite{proctor_featuremetric_2025}. Here, however, we will focus on predicting the 6-qubit QFT's process fidelity using \textit{effective error rates}.

The effective error rate and its prediction for the full circuit's process fidelity is as follows. For $K$ subcircuits of shape $(w,d)$, with measured process fidelities $\{ F_{w,d,i} \}_i=1^K$, the associated effective error rate is~\cite{seritan_benchmarking_2025}
\begin{equation}
    \epsilon_{w,d} = 1 -  \left(\prod_{i=1}^K F_{w,d,i} \right)^{1/(wdK)}.
\end{equation}
This error rate is then used to predict the fidelity of the full circuit as
\begin{equation}
    F = (1 - \epsilon_{w,d})^{w_c,d_c},
\end{equation}
where $w_c$ and $d_c$ are the width and depth of the full circuit, respectively.

In the left and central plots of Fig.~\ref{fig:subcirc-pred} we show the effective error rates and predicted full circuit fidelity for the subcircuit benchmarks of each circuit shape, as well as for the varied-size QFT circuits, as a function of circuit depth. For the simulated data (upper row), we observe that the subcircuits of each shape (colored lines) accurately predict the full circuit's fidelity (red dashed lines). In contrast, the predictions of the varied-size QFT circuits (black lines) do not accurately the full circuit's fidelity, for the simulated data, but with decreasing prediction error as the input size of the QFT circuit increases. However, for the experimental data (lower row), we observe the opposite effect: the subcircuits inaccurately estimate the full circuit's fidelity---with improving accuracy as the circuits get deeper---and the varied-size QFT circuits accurately predict $F$. This is, to our knowledge, the first comparison of these two approaches to predicting a circuit's fidelity from smaller circuits. This comparison was enabled by \oursoftware's flexible interface, and so we anticipate that it will enable more detailed studies of these---and other---benchmarking methods, and a better understanding of the merits and regime of reliability for different, complementary benchmarking methods.

\section{Discussion}
\label{sec:discussion}
In this paper we introduced \oursoftware, which is a tool for creating benchmarks from user-provided algorithms or circuits. We showed how \oursoftware~can be used to create efficient and robust benchmarks, and how these benchmarks can enable explorations of different aspects of the performance of contemporary quantum computer. Our demonstrations of \oursoftware~used circuits defined on physical qubits (i.e., NISQ computations), but \oursoftware~could also be applied to circuits defined on logical qubits protected by quantum error correction, in fault-tolerant quantum computing (FTQC) architectures. We note, however, that FTQC and NISQ architectures have many important differences, and reliable benchmarking of FTQC systems might require new methods to be developed and added to \oursoftware.

We showed how \oursoftware~can be used to improve existing benchmarks---to make them scalable and robust---and we demonstrated this with exemplar benchmarks from the QED-C's suite. At its core, \oursoftware~contains a protocol for efficient estimation of circuit process fidelities, and we showcased how that can be used to estimate different aspects of the performance of an algorithm's implementation---including the degradation in process fidelity due only to noise, the combined effect of noise and algorithmic approximations, and the in-situ impact of compiler optimizations in the presence of noise. Due to \oursoftware's simple interface and broad applicability, \oursoftware~could become an important component in benchmark development, handling many of the technical aspects of robust and scalable benchmark design while leaving the selection of interesting candidate circuits and algorithms to the user.

\section*{Code Availability}
\begin{small}
\oursoftware~is available within \texttt{pyGSTi} (version 0.9.14) and can be found at \text{https://github.com/sandialabs/pyGSTi}. All the benchmarks created in this work can be reproduced using \oursoftware~together with \texttt{qiskit}, \texttt{HamLib}, or the QED-C's open-source benchmarking suite, which can be found at 
\href{https://github.com/SRI-International/QC-App-Oriented-Benchmarks}{https://github.com/SRI-International/QC-App-Oriented-Benchmarks}.

\section*{Acknowledgements}
We thank Andrew Baczewski for helpful comments on the manuscript. N.S. thanks Aidan Wilber-Gauthier for helpful discussions. This material is based upon work supported by the U.S. Department of Energy, Office of Science (DE-FOA-0002253), National Quantum Information Science Research Centers, Quantum Systems Accelerator. T.P. acknowledges support from an Office of Advanced Scientific Computing Research Early Career Award. Sandia National Laboratories is a multi-program laboratory managed and operated by National Technology and Engineering Solutions of Sandia, LLC., a wholly owned subsidiary of Honeywell International, Inc., for the U.S. Department of Energy's National Nuclear Security Administration under contract DE-NA-0003525. All statements of fact, opinion, or conclusions contained herein are those of the authors and should not be construed as representing the official views or policies of the U.S. Department of Energy or the U.S. Government. This research used IBM Quantum resources of the Air Force Research Laboratory. The views expressed are those of the authors and do not reflect the official policy or position of IBM or the IBM Quantum team. The Quantum Economic Development Consortium (QED-C), comprised of industry, government, and academic institutions with NIST support, formed a Technical Advisory Committee (TAC) to assess quantum technology standards and promote economic growth through standardization. The Standards TAC developed the Application-Oriented Performance Benchmarks for Quantum Computing as an open-source initiative with contributions from multiple QED-C quantum computing members. Funding for N.P. was provided by Unitary Foundation and Quantum Computing Data.

\bibliographystyle{IEEEtran}
\bibliography{bibliography.bib,siekierski.bib}

\appendix

\subsection{QED-C Application-Oriented Benchmarking Suite}
\label{appen:qed-c}
In this appendix we review the QED-C's application-oriented benchmarking suite and further discuss how it can be interfaced with \oursoftware~to enable efficient and scalable benchmarks, which we demonstrate with examples in the main text. The QED-C's application oriented benchmarking suite is an evolving set of over 20 different application-based benchmarks for quantum computers, including algorithms such as phase estimation, Hamiltonian simulation, and the QFT. QED-C suite measures execution quality, runtime costs, and resource requirements for both single-circuit executions and iterative algorithms, using normalized classical (also called Hellinger) fidelity (defined in~\eqref{eq:def-classical-fidelity} in the main text) to assess circuit quality under noise. The first implementation of this benchmarking suite integrated all aspects of the benchmarking workflow, including algorithm definition, benchmarking circuit generation, circuit execution, and data analysis. Recently, however, the framework was modularized to enable more flexible interfacing with external benchmarking tools \cite{Patel2025-uy}, and we leverage this modularization in this work. The updated design separates the workflow into independent stages---problem generation, execution, and analysis---that can be accessed and run independently, as shown in the left column of Fig.~\ref{fig:qed_c_intergration}. This modular workflow enables interfacing the QED-C's suite with \oursoftware. In particular, the circuits defining a QED-C benchmark can be extracted (using a \textit{get\_circuits} flag) and then input into~\oursoftware, as shown in Fig.~\ref{fig:qed_c_intergration}

The purpose of interfacing the QED-C benchmarking suite with \oursoftware~is to enable the conversion of the QED-C's benchmarks into more robust and scalable benchmarks that measure process fidelity. Some of the benchmarks in the QED-C suite use specialized methods to enable them to scale to many qubits, such as picking particular inputs for a circuit that enable efficient classical computation of the expected output. These choices do not always result in a benchmark that will be predictive of the performance of that circuit on other input states. Conversely, some of the QED-C benchmarks are not scalable to many qubits, because they rely on exponentially-scaling classical computations. Interfacing the QED-C suite with \oursoftware~solves these problems, offloading this aspect of benchmark design to an independent, general-purpose tool.

\begin{figure}[t!]
\centering
\includegraphics[width=0.7\linewidth]{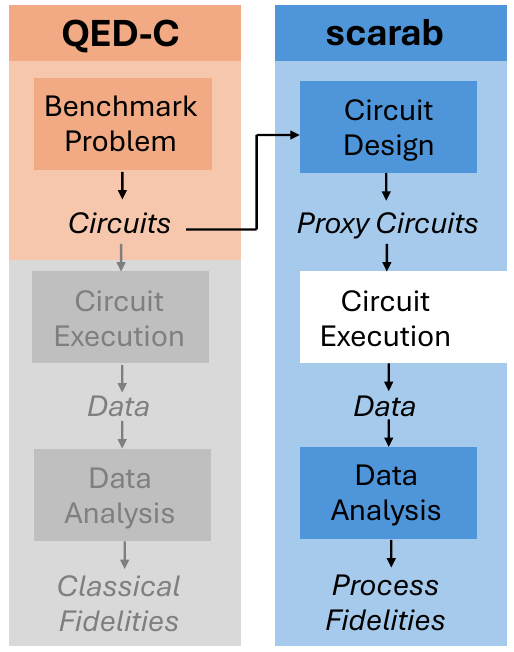}
\caption{\begin{small}\textbf{Interfacing the QED-C's suite with \oursoftware.} The modular QED-C benchmarking suite (left column) can be interfaced with \oursoftware~(right column) to transform any of the QED-C's benchmarks into efficient, scalable, and robust benchmarks that measure process fidelity. To do so, we can extract the circuits that are used to create a QED-C benchmark and instead input them into~\oursoftware. Similar interactions with other libraries of computational problems or quantum circuits are straightforward, and some of other examples of these interactions are demonstrated herein. \end{small}}
\label{fig:qed_c_intergration}
\end{figure}

\subsection{Hamiltonians}
\label{appen:hamiltonians}
In this appendix, we provide more information on the Hamiltonians obtained from HamLib~\cite{sawaya_hamlib_2024} that we simulate in Section~\ref{sec:ham-sim}.These are the transverse field Ising model (TFIM), Heisenberg, Bose-Hubbard, Fermi-Hubbard, and Max3SAT Hamiltonians. The TFIM Hamiltonian is given by
\begin{equation}
    H_{\text{TFIM}} = \sum_i h_i X_i + \sum_{\langle i,j \rangle} Z_i Z_j,
\end{equation}
where the second sum is over the edges $\langle i,j \rangle$ of the lattice. For our simulations, we consider a 1D lattice with periodic boundary conditions and $h_i = 2$ for all $i$.

The Heisenberg Hamiltonian is given by
\begin{equation}
    H_{\text{Heis}} = \sum_{i=1}^N X_i X_{i+1} + Y_i Y_{i+1} + Z_i Z_{i+1} + h_i Z_i.
\end{equation}
For our simulations, we consider a 1D lattice with periodic boundary conditions and $h_i = 2$ for all $i$.

The Fermi-Hubbard~\cite{hubbard_electron_1963} Hamiltonian is given by
\begin{equation}
    H_{\text{FH}} = -t \sum_{\langle i,j \rangle, \sigma} \left(c^\dagger_{i,\sigma} c_{j,\sigma} + c^\dagger_{j,\sigma} c_{i,\sigma}\right) + U \sum_{i} n_{i\uparrow} n_{i\downarrow},
\end{equation}
where $\langle i,j \rangle$ are lattice edges, $\sigma \in \{ \uparrow, \downarrow \}$ labels the fermion spin, $c$ and $c^\dagger$ are the fermionic creation and annihilation operators, respectively, and $n_{j\sigma} = c^\dagger_{j \sigma} c_{j \sigma}$ is the number operator associated with spin $\sigma$ and site $j$. The prefactor $t$ on the first sum is the tunneling strength, and $U$ is the on-site interaction strength. We set $t=1$, $U=12$ on a 1D lattice with periodic boundary conditions and a Brayvi-Kitaev~\cite{cao_quantum_2019} encoding.

The Bose-Hubbard Hamiltonian is given by
\begin{equation}
    H_{\text{BH}} = -t \sum_i \left( b^\dagger_{i+1} b_i + b^\dagger_i b_{i+1} \right) + \frac{U}{2} \sum_i n_i \left(n_i - 1\right),
\end{equation}
where $b^\dagger_i$ and $b_i$ are creation and annihilation operators respectively, $n_i = b^\dagger_i b_i$ is the number operator, $t$ is the tunneling strength, and $U$ is the site energy. For our simulations, we consider a 1D lattice with non-periodic boundary conditions that uses the Gray encoding, with $t=1$ and $U=10$.

The Max3SAT Hamiltonian is a sum of terms that correspond to 3-variable clauses. Each clause has the form
\begin{equation}
    \left( \neg \right)^{s_i} x_i \lor \left( \neg \right)^{s_j} x_j \lor \left( \neg \right)^{s_j} x_j,
\end{equation}
where $s_j$ = 1 if $x_j$ is negated in the clause and equals 0 otherwise. The corresponding term in the Hamiltonian is
\begin{equation}
    I - \frac{1}{8} \left[I + (-1)^{s_i} Z_i\right]\left[I + (-1)^{s_j} Z_j\right] \left[I + (-1)^{s_k} Z_k\right].
\end{equation}
The total number of clauses is given by $rn$, where $r$ is the clause ratio and $n$ is the number of qubits. We set $r = 2$ for our simulations and set \texttt{rinst}, the random instance flag used by HamLib to select a set of clauses, to \texttt{02}.
\end{small}
\end{document}